\documentclass[runningheads]{llncs}
\usepackage[T1]{fontenc}
\usepackage{amsmath}
\usepackage{amssymb}

\usepackage{tikz}
\usetikzlibrary{shapes.geometric, arrows.meta, chains, positioning}
\usepackage[colorlinks,linkcolor=red,anchorcolor=red,citecolor=blue]{hyperref}

\usepackage[ruled,linesnumbered]{algorithm2e}
\usepackage{graphicx}
\usepackage{subcaption}
\graphicspath{{./figs/}} 

\begin{document}
\title{AONA: A Comprehensive Architecture and Workflow Design for Global Agentic Collaboration}
\titlerunning{AONA architecture for Internet of Agents} 

\author{Jinliang Xu\thanks{Corresponding author.} 
\and
Runkai Zhu
\and
Bingqi Li
\and
Fanjie Nie
\and 
Jin Li
\and 
Jiagui Xie
}

\institute{China Academy of Information and Communications Technology, \\ No.52, North Huayuan Road, Haidian District, Beijing 100191, China \\
\email{\{xujinliang,zhurunkai,libingqi,niefanjie,lijin1,xiejiagui\}@caict.ac.cn}
}
\maketitle              
\setcounter{footnote}{0}
\begin{abstract}
The rapid advancement of Large Language Models (LLMs) has established autonomous agents as the core vehicles for artificial intelligence applications. 
However, existing Internet infrastructures, primarily relying on TCP/IP and DNS, are designed for human-centric, host-to-host data transmission, inherently lacking the semantic awareness, dynamic capability discovery, and decentralized trust mechanisms required for autonomous agent interactions. 
To address these limitations and break the closed ecosystems of single vendors, this paper proposes AONA (Agentic Overlay Network Architecture), a novel overlay network architecture for the Internet of Agents (IoA). 
We first provide a multi-disciplinary scientific defense for multi-agent collaboration, demonstrating its theoretical necessity over single super-intelligence through the lenses of organizational economics, scaling principles, and the Price of Anarchy.
AONA is then structured as a four-layer logical blueprint comprising the Base, Interconnection, Collaboration, and Application layers, which facilitates cross-protocol and cross-platform interoperability without disrupting the underlying physical network. 
To physically instantiate this blueprint, we design a distributed node infrastructure anchored by Management Root Nodes, Registry Service Nodes, Discovery Service Nodes, and Enterprise Intelligent Service Hubs for private domain integration. 
Finally, we detail the dynamic operational workflows-including zero-trust identity issuance, globally coordinated semantic taxonomy synchronization, intent-driven semantic discovery, and trusted metering for commercial settlement-that drive the network. 
This comprehensive architecture provides a robust, scalable, and secure foundation for the future of global agentic collaboration.

\keywords{Internet of Agents (IoA) \and
Multi-Agent Collaboration \and
Overlay Network \and
Semantic Service Discovery \and
Decentralized Identity (DID)\and
Large Language Models (LLM)\and
Distributed Infrastructure.}
\end{abstract}

\section{Introduction}
\begin{figure}[!ht]
\centering
\includegraphics[width=\textwidth]{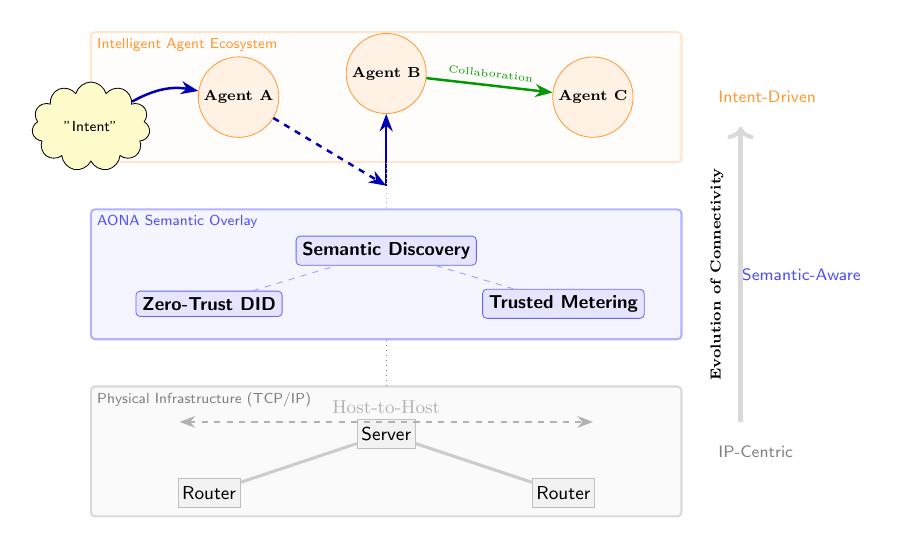}
\caption{The conceptual shift from traditional host-to-host connectivity to intent-driven agentic collaboration. AONA functions as a semantic overlay that leverages decentralized trust and discovery mechanisms to enable global multi-agent interaction without replacing the legacy network infrastructure.} \label{fig:whyAONA}
\end{figure}

The breakthrough of Large Language Models (LLMs) in 2023 has profoundly catalyzed the development of artificial intelligence, establishing autonomous agents as the core vehicles for large-scale AI applications. Unlike traditional programs that passively respond based on predefined rules, modern agents act as autonomous systems. Powered by LLMs as their \textit{brains} to understand human intent and conduct autonomous planning, and equipped with software or physical tools as their \textit{hands}, agents can execute complex tasks independently. To better illustrate this paradigm shift from traditional host-centric connectivity to intent-driven semantic collaboration, Fig. \ref{fig:whyAONA} presents a conceptual comparison of the legacy network and the AONA overlay.

However, as task scenarios become increasingly intricate, the limitations of single-agent systems in computational efficiency and resource constraints become apparent. Both theoretical proofs and recent engineering studies have demonstrated that multi-agent systems outperform single agents in handling complex, long-chain environments \cite{li2025acps,sharma2025collaborative}. By establishing a professional division of labor and team collaboration, agents can achieve complementary advantages and real-time error correction, analogous to human expert teams. This necessitates a fundamental paradigm shift from isolated intelligent tools to a globally collaborative network, paving the way for the Internet of Agents (IoA).

Realizing global-scale agentic collaboration introduces extreme system complexities. These include interaction complexity driven by autonomous and unpredictable high-concurrency traffic, architectural complexity in decentralized environments, and governance complexity where traditional centralized control fails to regulate highly autonomous entities. Existing Internet infrastructures, primarily built upon the TCP/IP protocol stack and the DNS, are inherently inadequate for these demands \cite{raskar2025beyond,chang2025agent,fleming2025layered}. Traditional networks are designed for host-to-host and human-centric data transmission. Specifically, DNS is not tailored for agentic environments: its update frequency is too slow to meet the sub-second requirements of dynamic agents; its trust model only verifies domain ownership rather than agent capabilities or safety; its metadata structure is too rudimentary to carry complex, natural-language capability descriptions; and its performance cannot scale to handle billions of autonomous updates and connections \cite{raskar2025beyond}.

Recognizing these critical bottlenecks, a strong consensus is rapidly forming across academia and industry regarding the necessity of a dedicated agentic network. Various conceptual ecosystems, such as the IoA, Agentic Internet, and Agentic Web, have been proposed, alongside specific functional protocols like Agent Name Service (ANS)\cite{cui2025agentdns,huang2026agent,cui2025soda,raskar2025upgrade,huang2025novel,yang2025survey}. The shared vision is to construct an evolutionary overlay network that preserves the existing physical infrastructure while providing exclusive, secure, and semantic-aware interoperability for agents across different platforms and organizations \cite{xu2026darwinnet}.

To systematically address these challenges and break the silos of single-vendor ecosystems, this paper proposes a comprehensive and standardized architecture for the Internet of Agents, i.e., AONA (Agentic Overlay Network Architecture). 
We first establish the scientific justification for multi-agent collaboration over single-agent systems from the perspectives of economics, complexity, and systems resilience.
We then design a novel four-layer logical blueprint comprising the Base, Interconnection, Collaboration, and Application layers. To physically instantiate this blueprint in a multi-stakeholder environment, we introduce a distributed node infrastructure anchored by management root nodes, registry service nodes, and discovery service nodes. Furthermore, we detail the dynamic operational workflows that drive the network, encompassing zero-trust identity issuance, global semantic discovery, and trusted commercial metering. Ultimately, this paper provides a robust technical blueprint for realizing secure, efficient, and scalable global agentic collaboration.

The remainder of this paper is organized as follows. Section \ref{sec:relatedworks} reviews the related works, discussing the current explorations and the limitations of existing frameworks in both academia and industry. 
Section \ref{sec:rationality} provides a multi-disciplinary scientific defense of multi-agent systems, demonstrating the theoretical necessity of decentralized collaboration.Section \ref{sec:fourlayer} introduces the logical blueprint of the AONA, detailing the four-layer architecture that enables cross-protocol and cross-platform interoperability. Section \ref{sec:physical} presents the physical infrastructure of the AONA, explaining the distributed node design which includes management root nodes, registry service nodes, and discovery service nodes. Section \ref{sec:workflow} elaborates on the dynamic operations of the network, providing comprehensive workflows for zero-trust identity issuance, global semantic discovery, and trusted commercial metering. Section \ref{sec:characteristics} analyzes the key characteristics and system complexities of the AONA, highlighting its evolutionary continuity from traditional networks alongside its revolutionary innovations in agentic interactions. Section \ref{sec:standardization} outlines the standardization and ecosystem governance framework, defining the core technical rules required for secure interoperability and fair value distribution. Section \ref{sec:roadmap} presents the implementation roadmap and the demonstration network, detailing the multi-phase deployment and testing strategies to validate the proposed architecture. Finally, Section \ref{sec:conclusion} concludes the paper and outlines potential directions for future research.

\section{Related Works}\label{sec:relatedworks}

\subsection{Enhanced Discovery and Indexing Architectures}

This category of work recognizes the fundamental limitations of the DNS in dynamic, agentic environments and proposes new indexing systems for agent discovery, identity resolution, and metadata verification.

\textbf{Project NANDA and the NANDA Index} \cite{raskar2025beyond,wang2025using} is a prominent proposal that envisions a lean, global index for the Internet of AI Agents. It decouples stable agent identifiers from dynamic metadata (AgentFacts), supporting sub-second updates, privacy-preserving lookups, and cryptographically verifiable capability assertions. Its design goals include lightweight indexing, multi-modal registration, and flexible routing, addressing DNS's slow update cycles and ownership-only trust model.

ANS \cite{huang2026agent} follows a similar vision, proposing a universal directory for secure AI agent discovery and interoperability. It extends beyond traditional DNS by incorporating semantic capability matching and verifiable identity features tailored for agentic AI.

While these works make significant strides in the resolution layer—mapping identities to verifiable metadata—they often lack a broader, system-wide blueprint. For instance, they typically do not explicitly define the physical distributed infrastructure, zero-trust execution workflows, or a comprehensive commercial metering system, leaving their concrete deployment and governance models in a multi-stakeholder environment largely undefined.

\subsection{Layered Protocol Stack Extensions}

A second line of research argues that the traditional TCP/IP or OSI model is insufficient for agent collaboration and proposes explicit new protocol layers to standardize agent communication and semantic alignment.

The \textbf{Layered Protocol Architecture for the Internet of Agents} \cite{fleming2025layered} explicitly proposes adding two new layers above HTTP/2/3 (which it terms the Application Transport Layer, L7): an Agent Communication Layer (L8) for standardizing message envelopes and interaction patterns, and an Agent Semantic Negotiation Layer (L9) for establishing shared context and semantic alignment before task execution.

\textbf{Agent Collaboration Protocols (ACPs)} \cite{li2025acps} represent a comprehensive protocol suite designed for the IoA, including sub-protocols for registration (ARP), discovery (ADP), interaction (AIP), and tooling (ATP). It aims to provide a unified framework to overcome the fragmentation of existing, scenario-specific protocols.

These proposals primarily focus on the protocol specification itself. However, they often fail to map these logical protocol requirements onto a concrete physical distributed infrastructure and do not elaborate on the dynamic operational workflows necessary to bring these protocols to life within a multi-domain commercial environment. They emphasize communication standards but may lack in system-level deployment models, cross-domain governance, and commercial operation mechanisms.

\subsection{Decentralized Trust and Identity Frameworks}

As agents become autonomous actors, establishing verifiable identity, accountability, and ethical governance becomes paramount. This category leverages decentralized technologies \cite{Jinliang2020Edgence} to create trust layers for agent ecosystems.

The \textbf{LOKA Protocol} \cite{ranjan2025loka} proposes a decentralized framework featuring a Universal Agent Identity Layer (UAIL) based on Decentralized Identifiers (DIDs) and Verifiable Credentials (VCs), intent-centric communication, and a Decentralized Ethical Consensus Protocol (DECP). It aims to embed identity, trust, and ethics into the protocol layer for responsible AI ecosystems.

The \textbf{Binding Agent ID (BAID)} framework \cite{lin2025binding} addresses the \textit{semantic gap} in traditional authentication by cryptographically binding a user's identity to the specific code an agent executes. It combines local biometrics, blockchain-based identity management, and zkVM-based code attestation to ensure accountability and credibility.

While these works make significant strides in decentralized identity and trust mechanisms, they are typically proposed as standalone trust protocols or frameworks. They often lack concrete plans for their integration into a larger, operational system focused on cross-domain collaboration. They may not explicitly detail how these trust mechanisms connect with practical functions like agent registration, and commercial metering, nor do they fully address balancing decentralization with operational manageability for global scale.

\subsection{Open Interoperability Protocols and Ecosystems}

This category encompasses industry-led initiatives to develop open standards and protocols that enable interoperability between agents and tools across different platforms, combating ecosystem fragmentation.

The \textbf{Model Context Protocol (MCP)} \cite{narajala2026enterprise,li2025netmcp} is a protocol for standardizing how applications and LLMs interact with external data sources and tools. It enables tools to expose a uniform interface, making them discoverable and usable by any MCP-compliant agent.

The \textbf{Agent-to-Agent (A2A) Protocol}, pioneered by Google \cite{a2a_spec_2024,a2a_announcement_2024}, provides a framework-agnostic standard for direct message exchange and delegation between agents. It focuses on enabling communication but leaves higher-level coordination and incentives to other layers\cite{yang2025survey}.

Note also the \textbf{Agent Network Protocol (ANP)} \cite{chang2025agent}, which proposes a comprehensive, AI-native three-layer protocol stack for the Agentic Web: an identity and encrypted-communication layer, a meta-protocol negotiation layer, and an application protocol layer (including concrete application-level specifications such as the Agent Description Protocol (ADP) and Agent Discovery mechanisms). ANP is designed to be minimally invasive to existing Internet infrastructure while supporting extensible, protocol-native agent interactions; in this sense it complements A2A’s point-to-point messaging by supplying a broader discovery, identity and negotiation substrate that can act as an interoperable companion or underlying layer for A2A-style exchanges.

The \textbf{Coral Protocol} \cite{georgio2025coral} positions itself as open infrastructure connecting the Internet of Agents. It builds upon MCP and integrates a blockchain-based payment layer to enable an incentivized marketplace of composable AI services, focusing on modular protocols and token-backed incentives.

 These protocols and initiatives primarily address specific aspects of agent communication and interoperability. However, they often focus on solving single-layer problems (e.g., communication formats or payments) and lack a holistic network infrastructure capable of connecting and uniformly routing these diverse protocols. They may not provide global semantic discovery mechanisms, or a broader multi-operator commercial settlement model needed for seamless interoperability of all these different protocol ecosystems at a global scale.

\subsection{Synthesis and Positioning of the AONA Architecture}

In summary, prior work has made significant strides in addressing specific challenges: indexing for discovery, new protocol layers for communication, decentralized systems for trust, and open standards for interoperability. The AONA architecture proposed in this paper synthesizes insights from all these categories into a unified, deployable whole. It provides a comprehensive logical blueprint (four-layer architecture), a physical instantiation model (distributed node infrastructure), and detailed dynamic workflows that tie together discovery, identity, trust, protocol adaptation, and commerce. Unlike proposals that focus on one layer or one function, AONA offers an integrated, system-level design for realizing a secure, scalable, and economically sustainable global agentic collaboration network.

\section{The Rationality of Multi-Agent Collaboration: Why MAS Outperforms Single Super-Intelligence}\label{sec:rationality}

The rapid evolution of LLMs has sparked a debate: can a single, all-knowing \textit{Super-Agent} eventually replace multi-agent systems (MAS)? While a centralized model offers simplicity, scientific analysis across multiple disciplines-including organizational economics, information theory, and systems engineering-demonstrates that MAS becomes necessary under bounded communication, cognition, and privacy constraints.

\subsection{Economic Optimality and Communication Constraints}
The scientific justification for MAS begins with Team Theory\cite{lemieux1972economic,simon1991bounded}. In an ideal environment without friction, a central controller with perfect information can make optimal decisions; however, this \textit{Revelation Principle} fails when communication is \textit{expensive} in terms of bandwidth, latency, or energy. 

\begin{itemize}
    \item \textbf{Control Loss vs. Communication Cost:} Scientific models prove that while delegating authority leads to a marginal \textit{control loss,} it significantly reduces the prohibitive costs of aggregating all local observations to a central node. 
    \item \textbf{Cognitive Limits:} Any single processor, regardless of its intelligence, faces \textit{information overload} in massive-scale environments. MAS resolves this by introducing organizational language and hierarchical task decomposition, effectively saving the system's \textit{cognitive budget}.
\end{itemize}

\subsection{Computational Complexity and Scaling Principles}
From the perspective of computational complexity, many real-world coordination problems are NP-hard. Centralized solvers face exponential growth in resource demands as problem variables increase\cite{modi2005adopt,kim2025towards}.

\begin{itemize}
    \item \textbf{Structure-Based Solving:} By utilizing the \textit{Interaction Graph} of a problem, MAS frameworks like Distributed Constraint Optimization (DCOP) can decompose global optimization into parallel, local sub-problems. This structure ensures that complexity grows relative to the task's \textit{treewidth} rather than the total number of nodes.
    \item \textbf{The Alignment Principle:} Quantitative research reveals a distinct \textit{Scaling Principle} for agents: in parallelizable tasks (e.g., multi-dimensional data retrieval or financial analysis), MAS provides a significant performance boost over single agents through \textit{context isolation}. This isolation prevents the reasoning noise and \textit{accuracy collapse} common in long-context single models.
\end{itemize}

\subsection{Emergence and System Resilience}
MAS exhibits \textit{Emergence}-where the collective intelligence exceeds the sum of individual parts \cite{reynolds1987flocks}. 

\begin{itemize}
    \item \textbf{The Reward Theorem:} Empirical and theoretical studies suggest that the expected cumulative reward of a communicating team is consistently higher than that of a non-communicating entity.
    \item \textbf{Fault Tolerance:} Unlike a single super-intelligence which represents a \textit{Single Point of Failure,} MAS is inherently robust. In distributed environments, the failure of an individual agent leads to graceful performance degradation rather than catastrophic system collapse.
    \item \textbf{Data Sovereignty:} MAS is a highly viable path for cross-organizational collaboration under strict privacy constraints. Through mechanisms like federated learning, agents can collaborate on sensitive data without exposing raw information, balancing collective intelligence with data sovereignty.
\end{itemize}

\subsection{Governance and the Price of Anarchy (PoA)}
In a world of autonomous entities, MAS provides a framework to manage conflicting incentives. PoA quantifies the efficiency loss caused by self-interested behavior, a concept originally formalized to measure the gap between social welfare and non-cooperative equilibria \cite{koutsoupias2009worst,xu2019blockchain}. In network-based tasks, this cost can reach significant multiples of the global optimum, particularly in scenarios with competitive resource routing\cite{roughgarden2002bad}.
It is defined as the ratio of the worst Nash equilibrium to the social optimum.
Without a collaborative protocol, PoA can lead to significant resource waste. The transition from \textit{self-interest} to \textit{Pareto efficiency} requires a standardized collaboration mechanism that aligns individual incentives with collective welfare.

In the decentralized governance of AONA, aligning the conflicting interests of diverse stakeholders is paramount. This can be addressed through multi-objective optimization techniques, as seen in path-selection mechanisms for vehicular networks that provide weakly Pareto-optimal solutions to balance network provider costs with user service quality, thereby minimizing the collective Price of Anarchy. Drawing parallels from mechanism design in crowdsourcing communities, where reward-penalty functions are utilized to align conflicting incentives among requesters, workers, and platforms \cite{xu2018reward}, AONA operationalizes these MAS principles through its Collaboration Layer to achieve Pareto efficiency and minimize the Price of Anarchy .

\subsection{Synthesis: AONA as the Realization of MAS Theory}
The theoretical imperatives mentioned above-communication efficiency, parallel scaling, fault tolerance, and incentive alignment-find their physical manifestation in the Agentic Overlay Network Architecture. By providing a semantic interconnection layer and a zero-trust collaboration framework, AONA functions as the \textit{nervous system} that allows these distributed intelligent entities to interact. It operationalizes the abstract MAS scientific principles into a deployable network infrastructure, facilitating the leap from isolated smart tools to a global, autonomous agentic ecosystem.

\section{The Logical Blueprint: Four-Layer Architecture of AONA}\label{sec:fourlayer}

\begin{figure}[!ht]
         \centering
         \includegraphics[width=\textwidth]{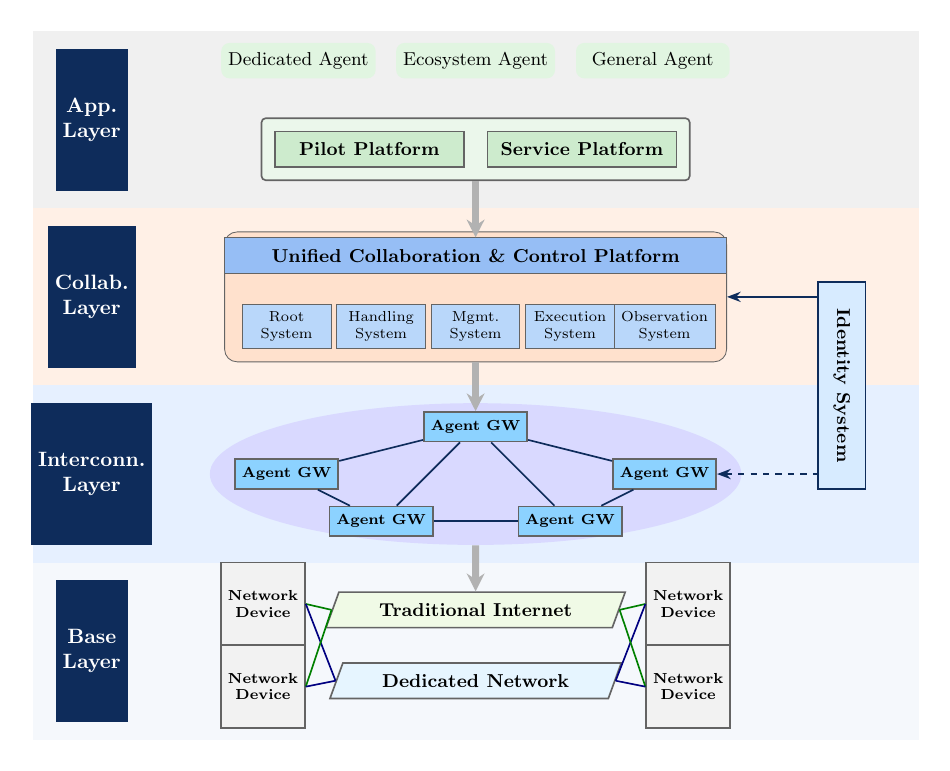}
         \caption{Physical networking and system topology.}
         \label{fig:network-framework}
     \end{figure}

\begin{figure}[!ht]
     \centering
         \includegraphics[width=\textwidth]{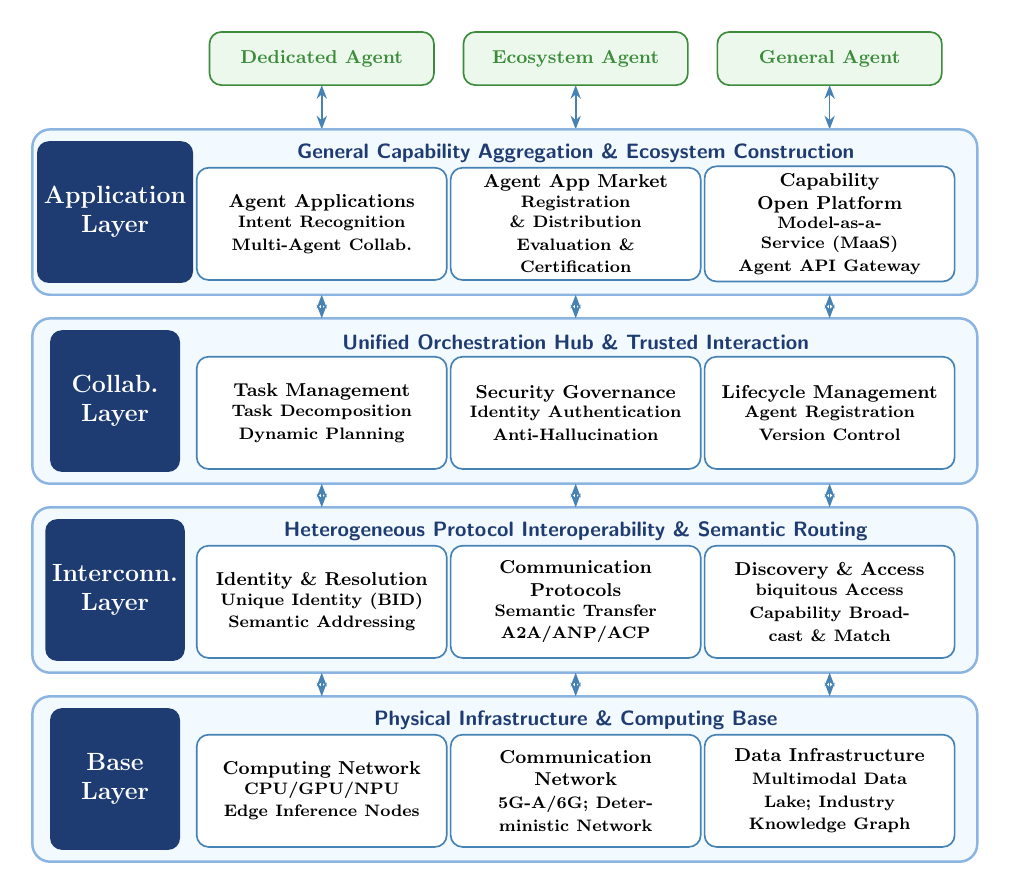}
         \caption{Four-layer technical functional stack.}
         \label{fig:tech-framework}
\end{figure}

The architecture of AONA is designed as an evolutionary overlay network built upon the existing TCP/IP infrastructure, leveraging IPv6 as its primary foundation for future network evolution while ensuring compatibility with current IP networks (including IPv4). It avoids the economically unfeasible and disruptive process of completely rebuilding the global network from the bottom up, instead achieving a capability leap through logical decoupling. As illustrated in the functional hierarchy (see Fig. \ref{fig:network-framework}), the AONA architecture is divided into four interrelated layers: the Base Layer, the Interconnection Layer, the Collaboration Layer, and the Application Layer. The comprehensive technical functional stack and its cross-layer dependencies are detailed in Fig. \ref{fig:tech-framework}.

\subsection{Base Layer: Infrastructure and Computing}
The Base Layer serves as the physical and computational foundation for the AONA, engineered to support high-frequency, multi-modal interactions among massive numbers of autonomous agents. It fully reuses and enhances traditional internet infrastructures, including mobile and satellite networks. To meet the stringent requirements of agentic traffic, this layer integrates Deterministic Networking (DetNet), Time-Sensitive Networking (TSN), and advanced IPv6-based mechanisms, such as SRv6, which represent an evolution of IP networking. These technologies guarantee microsecond-level low-latency and highly reliable transmission during complex, long-chain task executions. Furthermore, it provides ubiquitous computing clusters to satisfy the massive concurrent inference demands of LLMs, alongside native quantum security infrastructure to protect agent communications against future cryptographic threats.

\subsection{Interconnection Layer: Addressing and Protocol Adaptation}
Acting as the core communication hub, the Interconnection Layer shields the complexities of the underlying physical networks and provides a unified discovery and connection mechanism. It drives a fundamental paradigm shift in network routing: moving from finding a fixed IP address to finding a specific identity equipped with required capabilities. This layer utilizes Agent Gateways to provide cross-domain discovery and dynamic address resolution. Given that the agent ecosystem comprises highly heterogeneous protocols, this layer also embeds standardized protocol adaptation services. Converters for protocols such as the MCP, A2A, ANP, and AIP ensure seamless semantic and intent-level interoperability across different vendor platforms, effectively eliminating cross-organizational communication barriers.

\subsection{Collaboration Layer: Governance, Trust, and Orchestration}
The Collaboration Layer functions as the logical control brain of the AONA, designed specifically to manage the extreme governance complexity inherent in highly autonomous multi-agent systems. It enforces network-wide rules, maintains trust, and orchestrates tasks to resolve conflicts during cross-domain cooperation. This layer is composed of four critical subsystems. First, the Global Identity and Trust Root system establishes a Decentralized Verification under a Root-Assisted Trust Model, utilizing DID and VC for bidirectional authentication and dynamic authorization. Second, the globally coordinated semantic taxonomy system maintains a standardized capability label tree,
which provides a unified semantic reference for agent registration and discovery. Third, the Semantic Description Governance system standardizes capability dictionaries to maintain unified semantic contracts. Finally, the Collaboration Consensus system manages dynamic task decomposition, concurrent state synchronization, and trusted metering, providing the fundamental data for fair commercial billing and settlement across multiple network operators.

DIDs/VCs provide the foundational primitives for the AONA zero‑trust identity fabric. A DID is a portable, self‑controlled identifier whose resolution metadata (the DID Document) encodes public keys, verification methods and service endpoints; a VC is a cryptographically signed assertion issued by an attestor that binds claims (for example, issuer‑verified organizational attributes or capability attestations) to a subject DID. Together, DIDs and VCs enable cross‑domain trust without requiring a single central authority: parties can authenticate identities by resolving DID documents and verify assertions by validating VC signatures and issuance chains. This model supports fine‑grained, context‑dependent authorization (through scoped VCs and short‑lived delegation claims), rapid revocation and policy updates (via issuer revocation registries or revocation digests), and privacy‑preserving patterns (e.g., selective disclosure or holder‑bound credentials). By contrast, traditional PKI/DNS approaches rely on hierarchical, operator‑controlled namespaces and coarse ownership attestations that do not natively encode capability claims or support decentralized, cross‑operator authorization semantics; DNS lacks sub‑second, verifiable metadata updates and PKI certificates are typically coarse grained and centrally issued, making them less suitable for high‑frequency, multi‑stakeholder agent interactions. DIDs/VCs therefore better align with AONA’s requirements for decentralized, auditable, and fine‑grained identity and authorization across administrative domains.

Furthermore, AONA supports delegated authorization based on Verifiable
Credentials. Instead of exposing original identity credentials, agents can
issue scoped, task-specific credentials to counterparties. This enables
fine-grained, context-dependent access control while preserving privacy and
minimizing credential exposure.

\subsection{Application Layer: Intent-Driven Agent-as-a-Service}
At the top of the architecture, the Application Layer is designed for intent-driven value delivery, interacting directly with human users or top-level digital assistants. It encapsulates the complex underlying computing power, protocols, and coordination mechanisms into an accessible Agent-as-a-Service (AaaS) paradigm. By merely inputting natural language intents, users can trigger the entire closed-loop process of agent discovery, connection, and execution. This layer drives the large-scale application of AI-native services across diverse vertical industries-such as smart manufacturing, cross-border e-commerce, smart healthcare, and intelligent transportation-facilitating a structural leap from passive software operation to autonomous, end-to-end task fulfillment. 

\subsection{Protocol Stack Evolution: Compared with Traditional TCP/IP}
\begin{figure}
\centering
\includegraphics[width=\textwidth]{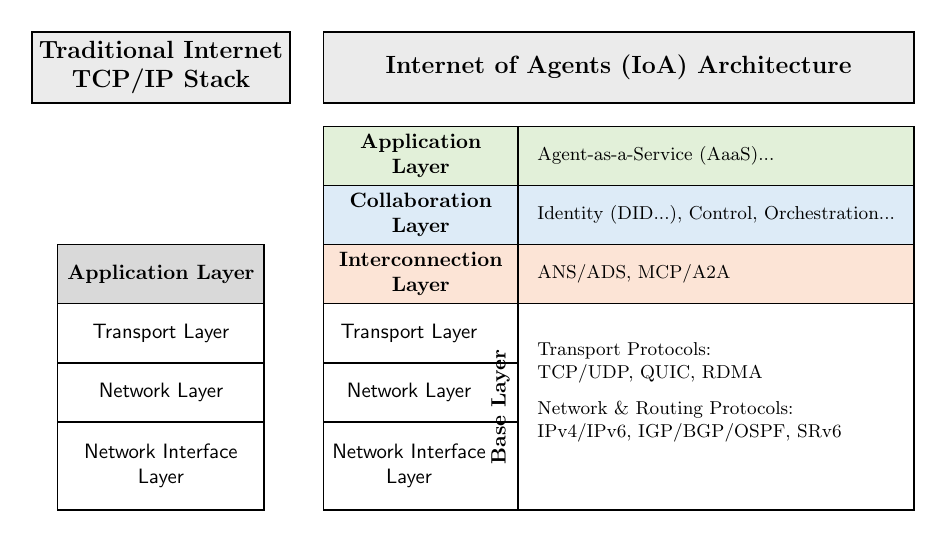}
\caption{Comparison of protocol stacks: Traditional TCP/IP model versus the AONA architecture.} \label{fig:layered-protocols}
\end{figure}

Unlike the traditional Internet, which relies on the standard TCP/IP four-layer protocol stack designed strictly for host-to-host data packet transmission, the AONA stack introduces a paradigm shift in network communication. As an overlay network, the primary architectural modifications and protocol innovations of the AONA occur \textit{above the traditional Transport Layer}, as compared in Fig. \ref{fig:layered-protocols}. 

In the AONA protocol stack, the Physical, Data Link, Network, and Transport layers are encapsulated within the \textbf{Base Layer}, which continues to utilize robust foundational protocols (such as DetNet, SRv6, and TCP/UDP) to ensure deterministic transmission. However, to fulfill the unique requirements of agent semantic understanding and dynamic discovery, the AONA superimposes the \textbf{Interconnection Layer} and the \textbf{Collaboration Layer} as entirely new logical tiers above the Transport Layer. These upper layers replace traditional application-agnostic data routing with agent-specific communication and governance protocols, such as the MCP, A2A protocol, and ATP. This structural evolution effectively upgrades the network from a static \textit{data-centric} transmission pipe to an autonomous \textit{intent-centric} collaboration ecosystem.

\subsection{Centralization and Decentralization: A Layered Interpretation}

AONA adopts a hybrid architectural paradigm that explicitly separates governance centralization from execution decentralization. 
This distinction is essential to avoid the conceptual ambiguity often associated with fully decentralized system claims.

At the \textbf{governance plane}, AONA introduces Management Root Nodes as a minimal coordination anchor. 
Their responsibilities are strictly limited to global policy definition, semantic taxonomy synchronization, and trust root establishment. 
This controlled centralization ensures global consistency, prevents semantic fragmentation, and enables interoperable trust across multiple administrative domains.

At the \textbf{execution plane}, however, AONA remains fully decentralized. 
All agent-to-agent interactions, task execution, and data exchanges are conducted in a peer-to-peer manner without reliance on centralized intermediaries. 
Discovery Service Nodes provide candidate recommendations but are not involved in authentication, routing enforcement, or execution, thereby preserving autonomy and scalability.

This layered separation aligns with established Internet design principles, where logically centralized coordination mechanisms (e.g., naming and governance) coexist with physically distributed execution. 
AONA therefore does not eliminate centralization entirely, but confines it to a narrowly scoped control layer while preserving decentralization in all runtime interactions.

\section{The Physical Infrastructure: Distributed Node Design}\label{sec:physical}

\begin{figure}[!ht]
\centering
\includegraphics[width=0.99\textwidth]{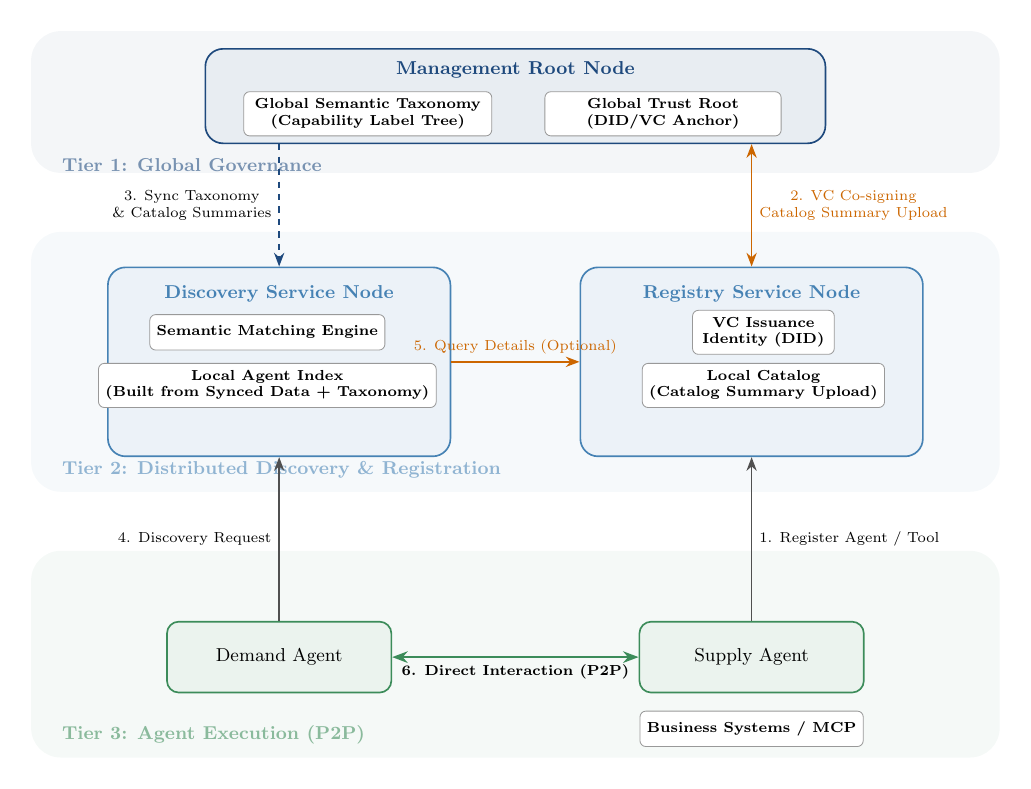}
\caption{Physical topology and interaction of the distributed node infrastructure. } \label{fig:service-network}
\end{figure}

To physically instantiate the four-layer logical blueprint without succumbing to the monopolistic bottlenecks of traditional centralized architectures, the AONA adopts a distributed node infrastructure. This architecture decomposes the core functionalities of the Interconnection and Collaboration layers and strategically deploys them across a multi-stakeholder ecosystem. By adhering to the principle of \textit{unified governance and distributed operations,} the physical topology is structured into a three-tier hierarchical architecture, as depicted in Fig. \ref{fig:service-network}: Tier 1 (Management Root Nodes) for global governance, Tier 2 (Service Exchange Nodes) comprising Registry and Discovery entities, and Tier 3 (Enterprise Intelligent Service Hubs) for private domain integration.

\subsection{Management Root Nodes: Global Governance and Trust Anchor}
At the apex of the AONA infrastructure are the Management Root Nodes.
Typically operated by authoritative industry regulators or neutral governance
bodies, these peer-to-peer interconnected nodes do not participate in runtime agent discovery, request routing, or business traffic forwarding. 
Instead, they operate strictly in the control plane, with responsibilities narrowly scoped to policy coordination, trust bootstrapping, and global semantic alignment.

To prevent governance overreach, Management Root Nodes are explicitly excluded from runtime decision-making, agent selection, and economic arbitration processes.

The Root Nodes fulfill two primary responsibilities, which is as shown in Fig. \ref{fig:service-network}. First, they act as the
Global Trust Root, issuing and managing authorization credentials (e.g.,
Verifiable Credentials) for Registry Service Nodes and Discovery Service Nodes,
thereby enabling these nodes to provide trusted registration and discovery
services to users and agents. Second, they function as a global data hub for
network-wide information synchronization. Registry Service Nodes periodically
report agent-related information to the Root Nodes, which then distribute this
data to Discovery Service Nodes across the network.

To ensure scalability, the Root Nodes may maintain a persistent backup archive
of agent information, but do not construct or maintain large in-memory,
fine-grained state tables for real-time query processing. Instead, the actual
semantic indexing and discovery operations are fully delegated to Discovery
Service Nodes. This design ensures that the Root Nodes remain lightweight,
while still providing global consistency and coordination capabilities.

Instead of maintaining a global directory of all agents, the Management Root Nodes
are responsible for defining and synchronizing a globally coordinated semantic taxonomy.
This taxonomy is structured as a hierarchical capability label tree, where each node
represents a standardized capability.

This semantic structure serves two primary purposes:
\begin{enumerate}
    \item  it provides a standardized reference for capability selection during agent registration,
    \item it enables coarse-grained filtering during discovery.
\end{enumerate}

The Root Nodes do not aggregate or store full agent metadata. Instead, Discovery Service Nodes
construct their own local directories based on locally available agent information and
the synchronized semantic taxonomy.

\subsection{Service Exchange Nodes: Edge Infrastructure and Routing}
Service Exchange Nodes form the backbone of the AONA's edge infrastructure. Distributed globally and operated by qualified third-party service providers (such as cloud computing platforms), these nodes directly interface with the agent ecosystem. Based on their functional roles, they are strictly categorized into two logical and physical entities: Registry Service Nodes and Discovery Service Nodes.

\subsubsection{Registry Service Nodes (Supply-Side Hubs)}
Registry Service Nodes act as the onboarding gateways for supply-side agents and MCP tools. Their primary responsibility is to handle enterprise qualification, service registration, and identity issuance. When an agent developer submits a registration request, this node conducts preliminary connectivity and security assessments. Upon successful verification, it generates a globally unique identifier  for the agent and requests the Root Node to co-sign a VC that securely binds the agent's identity to its capability metadata. Furthermore, Registry Nodes maintain local service catalogs and continuously monitor the online health status of registered agents, periodically synchronizing this operational data upward to the Root Node.

\subsubsection{Discovery Service Nodes (Demand-Side Hubs)}
Discovery Service Nodes (Demand-Side Hubs) Conversely, Discovery Service Nodes
serve demand-side agents by providing global semantic search and candidate
selection capabilities. Upon receiving globally synchronized agent information,
Discovery Nodes utilize LLM-based semantic matching and vector indexing to
identify potential target agents based on user intent.

Importantly, Discovery Service Nodes operate purely at the discovery layer.
They do not participate in authentication, protocol forwarding, or execution.
Instead, they return a set of candidate agents, which are treated as untrusted
results under a zero-trust assumption.

All subsequent identity verification and trust establishment are performed
directly between interacting agents. This design ensures that Discovery Nodes
remain lightweight and scalable, focusing solely on semantic discovery rather
than runtime interaction handling.

It is important to distinguish and clarify the operational boundary and cooperative relationship between Service Exchange Nodes (i.e., Registry Service Nodes and Discovery Service Nodes) and domain‑level Agent GWs. Service Exchange Nodes function as region/operator‑level exchange and directory hubs: they perform onboarding and qualification, DID/VC issuance, global catalog aggregation, and cross‑domain semantic search and routing decisions. In contrast, Agent GWs are deployed within enterprise or
local domains and serve as domain-level protocol adaptation components.
Their primary responsibility is to transform heterogeneous or legacy interfaces
(e.g., RESTful APIs, internal systems) into standardized agent-compatible
formats such as MCP. Agent GWs operate strictly within local domains, handling protocol conversion,
local policy enforcement, and system integration. They do not participate in
cross-domain discovery, authentication, or communication routing.

The strategic deployment of Registry and Discovery nodes is essential to alleviate backhaul network congestion, a critical bottleneck when agents exchange large-scale semantic metadata. By functioning as a trustless hybrid infrastructure, these edge nodes facilitate localized interaction between demand and supply agents, mirroring the efficiency gains observed in edge-enabled crowd-intelligence ecosystems \cite{xu2019blockchain}. All runtime discovery, routing, and interaction processes are handled by
Discovery Service Nodes, and do not involve the Management Root Node in the
data plane.

\subsection{Enterprise Intelligent Service Hubs: Private Domain Integration}
To integrate traditional industry ecosystems into the modern agentic network, the architecture introduces Enterprise Intelligent Service Hubs at Tier 3, which are specialized edge nodes deployed strictly within enterprise private intranets.

Many enterprises possess invaluable private digital assets, such as legacy databases (e.g., MySQL), industrial control software (e.g., SCADA, MES), and proprietary internal APIs, which are fundamentally incompatible with modern LLM interactions. The Enterprise Hub bridges this gap by encapsulating these heterogeneous data sources and converting them into standardized agentic protocols, such as the MCP. By proxying these encapsulated capabilities and hosting them on public Registry Service Nodes, the Enterprise Hub allows internal enterprise tools to be securely discovered and invoked by the global AONA network, thereby monetizing private digital assets without compromising data privacy or border security.

The design of Enterprise Intelligent Service Hubs aligns with the principles of privacy-preserving distributed learning \cite{sun2021edge}, where computational tasks are split between local front-ends and cloud-based back-ends. Similar to edge-enabled platforms that utilize asymmetric encryption and blockchain for stakeholder monitoring, AONA ensures that private digital assets are encapsulated locally and only semantic intents or sub-task results are exposed to the global registry service nodes, thereby eliminating raw data exposure risks.

Blockchain holds significant potential as a complementary infrastructure component for AONA, offering tamper-evident auditability and decentralized settlement capabilities that can materially strengthen trust and economic interoperability across operators. Despite the inherent strengths of blockchain and distributed ledger technologies in providing decentralized trust and tamper-evidence, it is impractical in a planetary-scale agent Internet to place every agent record or every agent-to-agent interaction fully on‑chain. With agent populations potentially reaching the order of $10^11$ and with very high interaction frequencies, naively on‑chain recording would impose untenable throughput, storage, latency and privacy costs. Therefore, we propose treating blockchain as an optional, targeted technology stack that is applied only where its properties yield clear benefits. Two representative usage patterns are particularly appropriate. First, blockchain can serve as an on‑chain bulletin board for governance and authorization events: Root/management nodes may publish compact, authenticated updates (for example, new authorizations, revocations, or policy digests concerning registry nodes, discovery nodes, or third‑party VC issuers) to the ledger; arbitrary network participants can then retrieve these authoritative updates via subscription, incremental polling, or event indexing, improving global consistency and verifiability without forcing large dynamic metadata onto the chain. Second, blockchain can act as a settlement and clearing substrate for economic interactions between agents: verifiable metering summaries, aggregated billing claims, or multi‑party payment commitments can be anchored or settled on‑chain to provide immutable evidence for cross‑operator clearing and dispute resolution. In practice we recommend a hybrid architecture: keep high‑frequency and sensitive metadata off‑chain (edge caches, secure logs, or privacy‑preserving databases) while anchoring succinct digests (e.g., Merkle roots) and governance/settlement artifacts on‑chain, thereby balancing performance, privacy and auditability.

\section{Dynamic Operations: Core Workflows for Agent Collaboration}\label{sec:workflow}

\begin{figure}[!ht]
    \centering
    \includegraphics[width=0.99\textwidth]{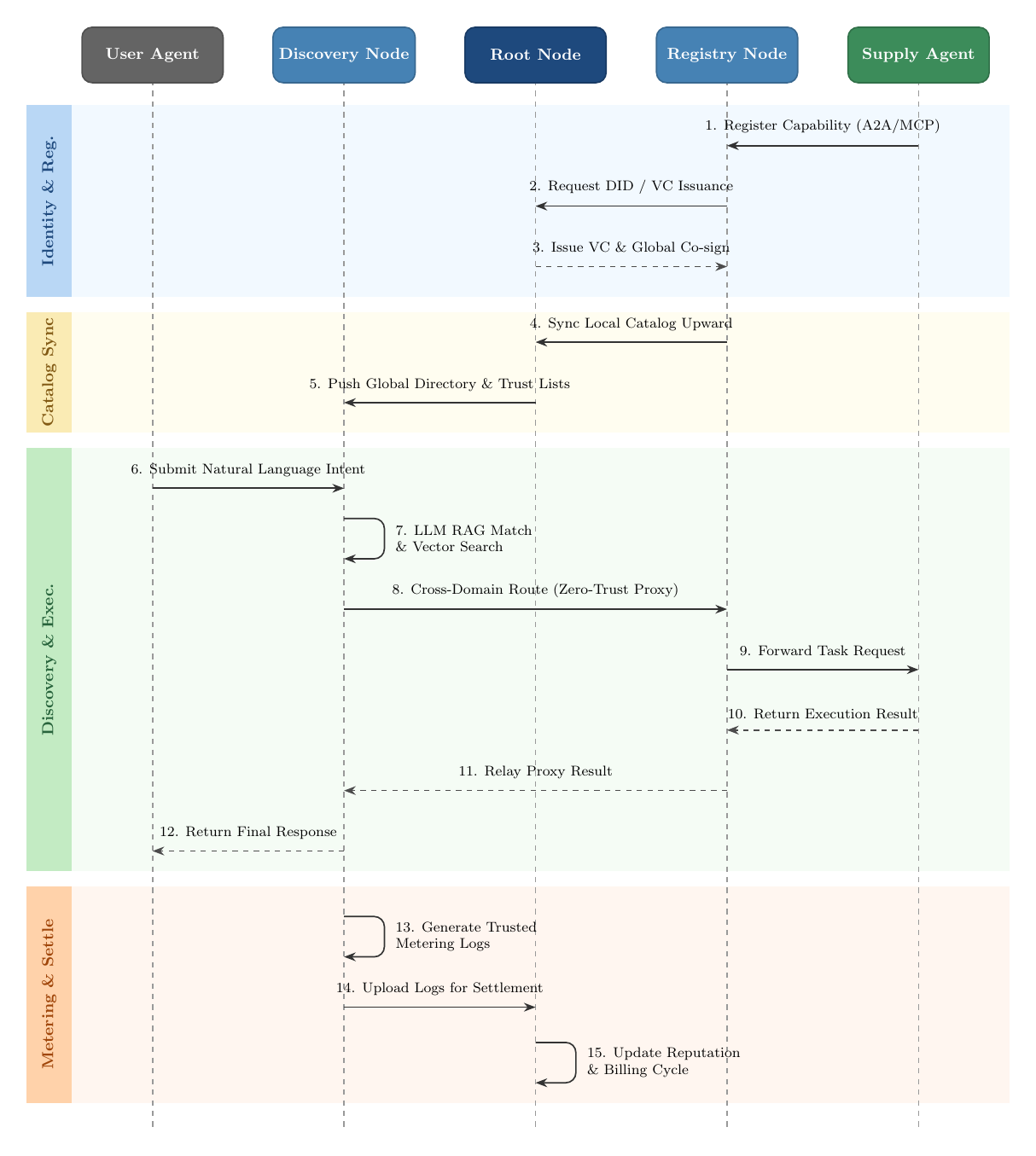}
    \caption{The full lifecycle sequence diagram of the IoA architecture, explicitly illustrating the core workflows corresponding to the four operational phases: (1) Zero-trust identity issuance and agent registration, (2) Global capability catalog synchronization, (3) Intent-driven discovery and (4) Trusted commercial metering and ecosystem settlement.}
    \label{fig:four_phase_sequence}
\end{figure}

The physical infrastructure and logical layers of the AONA are activated through a series of strictly orchestrated, closed-loop workflows. These dynamic operations dictate how intelligent entities are onboarded, discovered, authenticated, and compensated in a decentralized multi-operator environment. The system lifecycle is primarily driven by four core workflows: Agent Registration and Identity Issuance, globally coordinated semantic taxonomy synchronization and Semantic Discovery, and Trusted Commercial Metering. The interaction sequence of these phases is illustrated in Fig. \ref{fig:four_phase_sequence}.

\subsection{Agent Registration and Identity Issuance}
The onboarding workflow ensures that every agent or MCP tool entering the network possesses a verifiable identity and a standardized capability semantic contract. This process is primarily facilitated by the Registry Service Node acting as the supply-side gateway.

Initially, an agent developer or enterprise submits a registration request via a standardized SDK or developer portal. This submission includes the agent's basic information, natural language capability descriptions, skill tags, and supported transport protocols (which comprehensively cover STDIO for local processes, Server-Sent Events (SSE), and Streamable HTTP for remote streaming). To streamline the onboarding of legacy systems, the Registry Node supports the direct importation of OpenAPI (Swagger) specifications, utilizing a built-in parsing engine to automatically map traditional RESTful APIs into LLM-callable tool formats.

Upon receiving the configuration, the Registry Service Node conducts preliminary security assessments, including endpoint reachability tests and local command validation. Following successful validation, the node generates a unique DID for the agent. Crucially, to ensure global trust, the Registry Node applies to the Management Root Node to co-sign and issue a VC. This establishes a rigorous three-tier certificate chain: Root Certificate $\rightarrow$ Registry Node Certificate $\rightarrow$ Agent Certificate. 

Furthermore, to maintain backward compatibility with legacy systems, the architecture incorporates an Identity Adaptation mechanism. If an encapsulated MCP tool relies on traditional API Keys or Bearer Tokens, the Registry Node creates an internal mapping table that cryptographically binds these legacy tokens to the newly assigned DID. The generated DID document explicitly declares this authentication protocol, and the provided Identity Adapter SDK seamlessly translates DID-based challenge-response signatures into legacy authentication formats during runtime. Finally, the Registry Node initializes a heartbeat monitoring mechanism to continuously track the agent's online status and lifecycle events (e.g., updates, suspension, or deletion).

\subsection{globally coordinated semantic taxonomy synchronization and Semantic Discovery}
To break down information silos and enable intent-driven \textit{global agent routing,} the AONA employs a hierarchical directory synchronization and semantic discovery workflow.

The synchronization process operates bidirectionally. First, Registry Service Nodes compile their local databases of registered agents and MCP tools into lightweight catalog summaries. These summaries---containing service identifiers, capability classification codes, operational statuses, and endpoint URLs---are continuously uploaded to the Management Root Node through either periodic full-syncs or real-time incremental updates. The Root Node subsequently aggregates and resolves any semantic conflicts to form a unified Global Semantic Tree. This global directory tree is then pushed downward to all demand-side Discovery Service Nodes across the network, ensuring that every edge node possesses a consistent view of the global capability landscape.

When a demand-side entity, such as a user's personal digital assistant, requires assistance to complete a complex task (e.g., \textit{query cross-border e-commerce suppliers}), it submits a natural language intent to its connected Discovery Service Node. Instead of relying on exact string matching, the Discovery Node leverages an embedded LLM RAG engine and vector databases to perform multi-dimensional semantic matching against the synchronized global catalog. The engine evaluates capability descriptions, tool schemas, and operational health metrics to generate a Top-N recommendation of the most suitable target agents. If a local match is insufficient, the system relies on globally synchronized agent information available at Discovery Service Nodes to identify candidates
across different domains, without requiring the Root Node to participate in
runtime routing or query processing.

\subsection{Zero-Trust Authentication and Direct Execution}

Once a target agent is discovered, a secure communication channel is established directly
between the invoking agent (or its gateway) and the target agent.

AONA adopts a Zero-Trust model in which each interaction requires independent authentication.
Both parties perform mutual verification using DID resolution and VC validation,
ensuring that trust is established without relying on centralized intermediaries.

Discovery Service Nodes are not involved in the execution path.
Instead, they provide the necessary discovery results and metadata required
for establishing secure end-to-end connections.

This design avoids centralized bottlenecks and ensures scalability and autonomy
in agent interactions.

\subsection{Performance Scalability and Efficiency Optimization}
To address the inherent bottlenecks of traditional DNS extension schemes—specifically their reliance on static long-caching mechanisms, limited single-point processing capacity, and inability to adapt to millisecond-level agent creation or high-frequency dynamic updates—AONA achieves superior performance scalability through multi-dimensional architectural innovations. 
\begin{itemize}
    \item Network Distribution and Discovery: At the distribution level, the Management Root Node is designed to remain
lightweight by avoiding real-time query processing and fine-grained state
management. Although it serves as a global data coordination hub, it does not
maintain large in-memory semantic indices. Instead, it focuses on data
aggregation, backup archiving, and network-wide distribution, while delegating
all semantic discovery and matching tasks to Discovery Service Nodes.
    \item Coarse-to-Fine Matching: By  generating multi-level capability tag lists, the architecture enables Discovery Service Nodes to perform \textit{coarse-filtering pre-processing}. This significantly reduces the computational overhead required for subsequent fine-grained semantic matching, drastically enhancing retrieval performance for massive user intents.
    \item Decentralized Authentication: At the interaction level, the architecture bypasses cumbersome centralized redirection handshakes. By utilizing DIDs and  VCs, it facilitates millisecond-level, point-to-point local offline signature verification and concurrent request validation.
\end{itemize}      
This decentralized approach perfectly aligns with the stringent performance requirements for ultra-low latency and high-frequency interactions necessitated by a global ecosystem of trillions of agents.  

\section{Key Characteristics and System Complexities of AONA}\label{sec:characteristics}

The AONA represents a structural leap from a traditional \textit{Internet of Humans} to a globally collaborative \textit{Human-Machine-Intelligence} paradigm. Its core characteristics can be systematically analyzed through the dual lenses of evolutionary continuity and revolutionary innovation, alongside the inherent complexities of multi-agent environments.

\subsection{Evolutionary Continuity}
The AONA is not a disruptive replacement of the existing Internet, but rather a systematic evolution built upon its established strengths. This continuity is manifested in several dimensions:
\textbf{Architectural Continuity:} The AONA leverages the existing TCP/IP protocol stack, primarily building upon IPv6 infrastructure while retaining compatibility with current IP networks, thereby preserving the fundamental end-to-end packet switching design. By superimposing a logical \textit{Agent Collaboration Layer,} it ensures seamless migration for legacy devices and reuses mature encryption and access control mechanisms.
\textbf{Spirit and Ecosystem Continuity:} It inherits the open, decentralized ethos of the Internet, maintaining network simplicity while pushing intelligent complexities to the edge nodes. Furthermore, it extends the traditional \textit{human-to-human} and \textit{human-to-machine} connections into \textit{human-to-agent} and \textit{agent-to-agent} ecosystems, catalyzing the evolution of terminal devices into autonomous intelligent entities.
\textbf{Value Continuity:} The ultimate mission remains human-centric. Humans retain the roles of goal-setting and supervision, while agents act as partners to execute tasks, thereby upgrading traditional internet applications (e.g., communication, collaboration, commerce) into highly autonomous smart scenarios.

\subsection{Revolutionary Innovations}
Despite its continuous foundations, the AONA introduces fundamental paradigm shifts across the network lifecycle:
\textbf{New Subjects and Interaction Models:} Traditional networks connect passive hardware devices; in contrast, the AONA connects autonomous agents capable of initiating interactions, dynamically adapting to environments, and breaking the passive \textit{request-response} model. Users can issue high-level intents via multimodal inputs (text, voice, vision), and agents will autonomously decompose these intents, route the requests, and orchestrate network resources to achieve closed-loop execution.
\textbf{New Network Architecture:} The deployment topology shifts from a centralized client-server model to a distributed \textit{Cloud-Edge-Device} architecture. Because autonomous agents generate massive amounts of unpredictable, high-concurrency uplink traffic during edge-to-edge communications, the underlying transmission networks must be fundamentally optimized for bidirectional traffic balancing.
\textbf{New Governance Paradigm:} The AONA framework provides a novel paradigm for agentic name-spaces, complementing traditional DNS-based governance models like those managed by ICANN to meet the dynamic requirements of autonomous AI entities

\subsection{System Complexities}
The global deployment of AONA must address extreme system complexities. These include \textit{Subject Complexity} (heterogeneous agents with conflicting goals), \textit{Interaction Complexity} (unpredictable, high-concurrency traffic requiring semantic conflict resolution), and \textit{Governance Complexity} (the inadequacy of centralized control over highly autonomous entities). Overcoming these challenges necessitates a robust standardization framework and distributed ecosystem governance.

\section{Standardization and Ecosystem Governance}\label{sec:standardization}

\begin{figure}[!ht]
    \centering
    \includegraphics[width=0.99\textwidth]{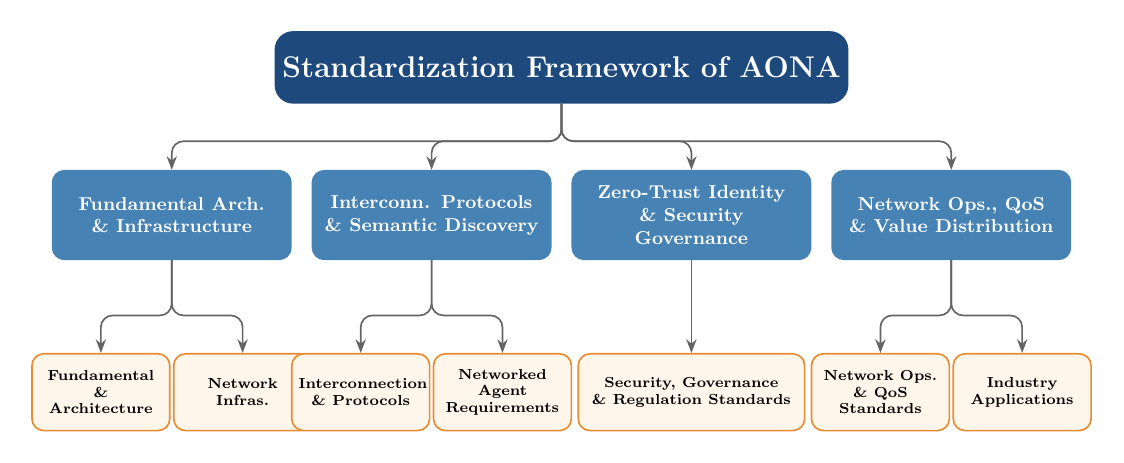}
    \caption{The hierarchical tree of the AONA standardization framework. The root architecture is supported by four core pillars, which are further decomposed into seven comprehensive standard sub-systems covering the full lifecycle from physical infrastructure to application governance.}
    \label{fig:standard-framework}
\end{figure}

To facilitate the large-scale deployment and high-quality development of AONA, a comprehensive standard system framework must be established. Overcoming the fragmentation of existing AI ecosystems requires a full-stack standardization loop that transitions the IoA from a conceptual model to a telecommunications-grade infrastructure. Building upon the seven-domain standard system framework-which encompasses foundational architecture, interconnection protocols, networked agent requirements, infrastructure, operations, industry applications, and security governance-this section details the four core pillars of the IoA standardization and ecosystem governance, the structure of which is visualized in Fig. \ref{fig:standard-framework}.

\subsection{Fundamental Architecture and Infrastructure Standards}
This pillar defines the top-level reference architecture and the physical evolution of the network. From a macro perspective, it establishes the unified terminology, conceptual models, and the strict mapping relationships between the logical application layers and the physical network infrastructure. 

To support the high-frequency semantic interactions of autonomous agents, this standard domain specifies the technical requirements for upgrading traditional network facilities. It mandates the deployment of specialized Agent Gateways and Agent Routers, and outlines the adaptation rules for IPv6+ and TSN to ensure deterministic transmission. At the implementation level, this is codified in the \textit{IoA General Technical Requirements}, which strictly defines the functional architecture, access management, and computing-network integration interfaces for the Registry Management Nodes and Discovery Service Nodes at the network edge. This prevents heterogeneous deployment architectures and guides operators in building standardized IoA infrastructure.

\subsection{Interconnection Protocols and Semantic Discovery Requirements}
The core of the IoA is the ability of agents to dynamically find and understand each other across different domains. This standardization pillar focuses on universal interconnection protocols, including cross-platform semantic interaction templates, heterogeneous agent interoperability requirements, and network access compliance. 

To solve the challenge of global agent addressing, this pillar incorporates the \textit{IoA Discovery Technical Requirements}. It standardizes the semantic discovery and routing mechanisms, defining how Discovery Service Nodes utilize local vector indexing for capability matching. Furthermore, it addresses the critical gap in cross-domain routing by standardizing the Relay routing protocols via Root Nodes when local discovery fails, alongside the data formats for globally coordinated semantic taxonomy synchronization and capability broadcasting. This ensures that agents can achieve \textit{register once, discoverable globally} without being locked into a single vendor's ecosystem.

\subsection{Zero-Trust Identity and Security Governance}
Security and trust are the prerequisites for autonomous agent collaboration. This standard pillar establishes a decentralized governance system designed to monitor, audit, and secure agent behaviors across the network. It defines the models for decentralized trust evaluation, full-lifecycle digital credential management, and cross-domain operation auditing to prevent malicious actions and mitigate cyber-attacks.

At the operational core is the \textit{IoA Identity Management Specification} [8]. It details a hierarchical zero-trust authentication mechanism based on DID and VC. This specification establishes a robust three-tier trust chain: Registry Nodes conduct initial verification of the agent's corporate entity, while the Global Root Node serves as the final authority to issue globally recognized VC credentials. Additionally, it standardizes millisecond-level anomaly recognition and blacklist synchronization protocols, ensuring that non-compliant agents are immediately isolated from the network.

\subsection{Network Operations, QoS, and Value Distribution}
For the IoA to thrive as a commercial ecosystem, it must ensure fair value distribution and telecommunications-grade service quality. This standard domain defines the Service Level Agreements (SLA), end-to-end Quality of Service (QoS) evaluation metrics \cite{xu2018reward}, and OpenAPI capability exposure norms. It also extends these operational standards to vertical industry applications, establishing cross-domain collaboration templates for smart manufacturing, connected vehicles, and digital government.

Crucially, to maintain market order and prevent platform monopolies, this pillar implements the \textit{IoA Operational Technical Requirements}. This standard resolves the complex issues of cross-operator billing and data ownership. It mandates that Discovery Nodes generate tamper-proof \textit{Trusted Metering Logs} for every routed task. Furthermore, it introduces a \textit{Contribution Tracing Protocol} which legally requires that the original contributor's DID be included in the return metadata whenever an agent utilizes a third-party tool or knowledge base. By standardizing electronic settlement receipts verified by the Root Node, this framework fundamentally eradicates \textit{data vampirism} and ensures fair commercial compensation for all AI developers.

\section{Implementation Roadmap and Demonstration Network}\label{sec:roadmap}

\begin{figure}[!ht]
\centering
\includegraphics[width=\textwidth]{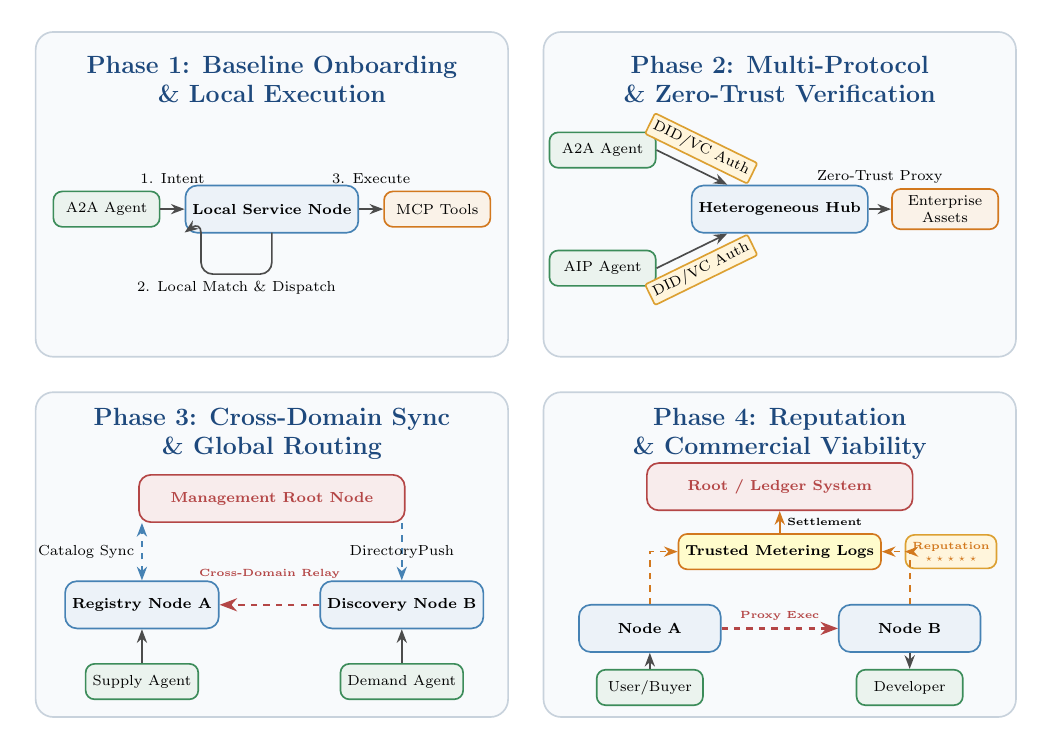}
\caption{The evolutionary roadmap of the AONA demonstration network, illustrating the progression from single-domain execution to a globally synchronized, commercially viable ecosystem.} \label{fig:four-phases}
\end{figure}

The theoretical architecture and standards of the AONA are being actively validated through a large-scale Demonstration Network. This network incorporates diverse stakeholders: authoritative regulators operating the Root Nodes, telecommunications and regional enterprises operating Registry Service Nodes, AI platform giants operating Discovery Service Nodes, and independent developers providing intelligent tools. The deployment and testing of this network follow a rigorous four-phase implementation roadmap, summarized in Fig. \ref{fig:four-phases}.

\textbf{Phase 1: Baseline Registration and Single-Domain Execution.} 
The initial phase validates the foundational onboarding and execution loops. It tests the successful registration of A2A-protocol agents and MCP tools into the edge nodes. The primary goal is to verify the closed-loop process of intent-driven discovery, task orchestration, and execution within a localized environment.

\textbf{Phase 2: Multi-Protocol Compatibility and Zero-Trust Verification.} 
The second phase expands the network's heterogeneity by integrating cross-protocol compatibility. It validates the registration and interoperability of agents utilizing diverse protocols, including A2A and AIP. Critical security tests are conducted to verify DID authentication and VC validation during agent-to-agent communications. Additionally, preliminary tests regarding the integration of underlying telecommunication networks are initiated.

\textbf{Phase 3: Cross-Domain Synchronization and Global Routing.} 
Phase three activates the distributed multi-node topology. It tests the seamless integration between edge Service Nodes (Registry and Discovery) and the Management Root Node. The focus is on validating the upward reporting of local catalogs from Registry Nodes, the aggregation of a unified global directory at the Root Node, and the subsequent downward distribution to Discovery Nodes. This phase verifies the network's capability to execute cross-domain agent discovery and relay routing.

\textbf{Phase 4: Optimization, Reputation, and Commercial Viability.} 
The final phase focuses on scaling performance and establishing ecosystem trust mechanisms. It aims to optimize the latency and semantic precision of the global discovery algorithms. Most importantly, it introduces and validates dynamic reputation and evaluation mechanisms for agents based on historical execution logs, laying the groundwork for a self-regulating, commercially viable global agent economy.

\section{Conclusion}\label{sec:conclusion}

From a systems perspective, AONA can be viewed as a hybrid distributed system that combines elements of hierarchical coordination and peer-to-peer execution, balancing global consistency with local autonomy. By integrating a four-layer logical blueprint, a distributed multi-stakeholder node infrastructure, and zero-trust operational workflows, this architecture successfully breaks single-vendor silos, enabling cross-protocol interoperability and trusted commercial metering without disrupting the underlying physical networks. While this design establishes a robust foundation for the agent economy, realizing a fully mature global intelligent network necessitates continuous exploration. Future research must focus on optimizing the latency and precision of semantic discovery algorithms to handle complex multi-step intent decomposition, as well as establishing dynamic, network-wide reputation and evaluation mechanisms based on immutable execution logs. Moreover, exploring deeper integration with next-generation physical infrastructures, such as 6G and low-earth orbit satellite networks, will be essential to support ubiquitous, high-bandwidth agent interactions. Finally, fostering international consensus and open standardization through global organizations remains the cornerstone for unifying semantic dictionaries and cross-border governance. Ultimately, the continuous evolution of the AONA will drive a structural leap in artificial intelligence, empowering a truly open, secure, and globally interconnected agent ecosystem.

\bibliographystyle{splncs04}
\bibliography{references}

\end{document}